\documentclass{aa}
\usepackage{natbib}
\usepackage{graphicx}
\usepackage{txfonts}
\usepackage{amssymb}

\def\lya{Ly$\alpha$}
\def\lyb{Ly$\beta$}

\begin{document}
   \title{Signatures of transition region explosive events in hydrogen \lyb~ profiles}
   \author{M. Zhang \inst{1}
          \and L.-D. Xia \inst{2, 3}
          \and H. Tian\inst{3, 4}
          \and Y. Chen\inst{2}}
   \institute{CAS Key Laboratory of Basic Plasma Physics, School of
    Earth and Space Sciences, Univ. of Science and Technology of China,
    Hefei, Anhui, China
   \and Shandong Provincial Key Laboratory of Optical Astronomy and Solar-Terrestrial Environment, School of Space Science and Physics, Shandong Univ. at Weihai,
    Weihai, Shandong, China\\
    \email{xld@sdu.edu.cn}
    \and Max-Planck-Institut f\"ur Sonnensystemforschung,
   Max-Planck-Str. 2, 37191 Katlenburg-Lindau, Germany
   \and School of Earth and Space Sciences, Peking University, China
   }

\date{Received 12 February 2010 / Accepted 10 May 2010}

\abstract
   {}
   {We search for signatures of transition region explosive events (EEs) in hydrogen \lyb~ profiles.}
   {Two rasters made by the SUMER (Solar Ultraviolet Measurements of Emitted Radiation) instrument on board SOHO
   in a quiet-Sun region and an equatorial coronal hole are selected for our study. Transition region
   explosive events are identified from profiles of C~{\sc{ii}}~1037{\AA} and O~{\sc{vi}}~1032~{\AA}, respectively.
   We compare \lyb~profiles during EEs
   with those averaged in the entire quiet-Sun and coronal-hole regions. The relationship between the peak
   emission of \lyb~profiles and the wing emission of C~{\sc{ii}} and O~{\sc{vi}} during EEs is investigated.
   }
   {We find that the central part of \lyb~ profiles becomes more reversed and the distance of the two peaks
   becomes larger during EEs, both in the coronal hole and in the quiet Sun. The average \lyb~ profile of the
   EEs detected by C~{\sc{ii}} has an obvious stronger blue peak. During EEs, there is a clear correlation between
   the increased peak emission of \lyb~profiles and the enhanced wing emission of the C~{\sc{ii}} and O~{\sc{vi}}
   lines. The correlation is more pronounced for the \lyb~ peaks and C~{\sc{ii}} wings, and less
   significant for the \lyb~ blue peak and O~{\sc{vi}} blue wing. We also find that the \lyb~ profiles are more reversed
   in the coronal hole than in the quiet Sun.
   }
   {We suggest that the jets produced by EEs emit Doppler-shifted \lyb~photons, causing enhanced emission at positions of
   the peaks of \lyb~profiles. The more-reversed \lyb~ profiles confirm the presence of a larger opacity in the coronal hole
   than in the quiet Sun. The finding that EEs modify the \lyb~line profile in QS and CHs implies that one should
   be careful in the modelling and interpretation of relevant observational data.}

   \keywords{Sun: transition region --
             Sun: UV radiation --
             Line: profile
            }
    \titlerunning{Signatures of EEs in \lyb~ profiles}
    \authorrunning{M. Zhang et al.}
   \maketitle

%
\section{Introduction}
Transition region (TR) explosive events (EEs) are small-scale
dynamic phenomena often observed in the far and extreme ultraviolet
(FUV/EUV) spectral lines formed in the solar transition region. They
were detected for the first time by the NRL/HRTS instrument
\citep{Brueckne83}. Since 1996, data obtained by the SUMER (Solar
Ultraviolet Measurements of Emitted Radiation) spectrograph
\citep{Wilhelm95,Wilhelm97} have been widely used to study EEs. With
high spatial and spectral resolution, and wide spectral coverage,
SUMER has greatly increased our knowledge of EEs. EEs are
characterized by non-Gaussian and broad profiles with enhancements
in the blue/red wings with an average line-of-sight Doppler
velocities of $\sim$100~km/s \citep{Dere89,Innes97a}. They have a
small spatial scale of about 1800 km and a short lifetime of about
60 s on average \citep{Teriaca04}. Explosive events tend to occur
along the boundaries of the magnetic network, where weak
mixed-polarity magnetic features are present
\citep{Porter91,Chae98,Teriaca04}. As EEs are often found to be
associated with magnetic cancelation and reveal bi-directional flows
with high velocities comparable to the local Alfv\'{e}n velocity,
they have been suggested to be a consequence of small-scale magnetic
reconnections \citep{Innes97b}. Sometimes EEs are found to burst
repeatedly in the same region, possibly a result of repetitive
reconnections triggered by P-mode oscillations or transverse
oscillations of the flux tubes \citep{Ning04,Doyle06,Chen06}.
Although EEs are best seen in typical TR lines, they can generally
be detected in spectral lines with formation temperatures ranging
from $\sim$$10^4$ to 5$\times10^5$ K
\citep{Madjarska02,Teriaca02,Popescu07}.

\begin{figure*}[t]
   \sidecaption
   \includegraphics[width=12.5cm]{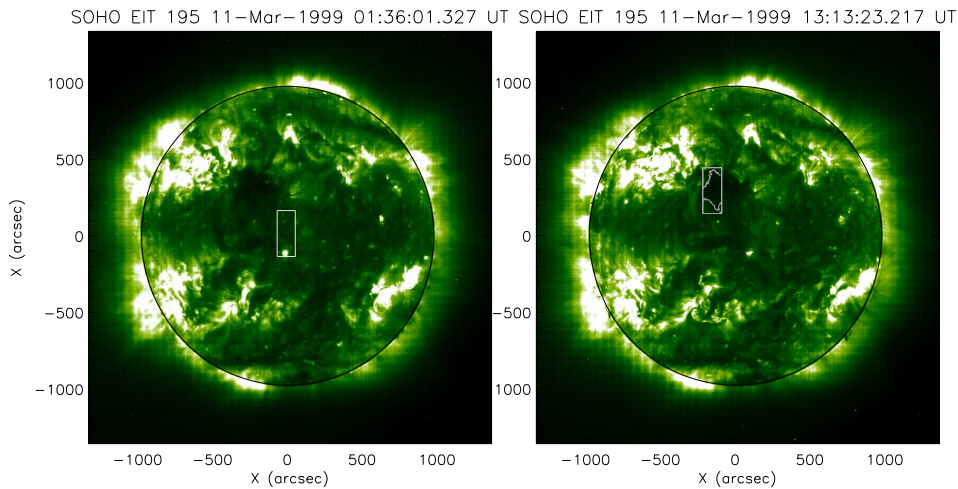}
   \caption{Two EIT images in the 195~{\AA} passband. The white rectangles
   indicate the scanned regions by SUMER (left: QS right: ECH), the curve on the right image outlines
    the ECH boundary.}\label{Fig_1}
\end{figure*}

\begin{table*}[t]
  \caption[]{Information of the SUMER observations}
  \label{tab1} \centering
  \begin{tabular}{ccccccc}
  \hline
Date & Time & Solar X & Solar Y & Detector & slit & Exposure time\\
\hline
1999.03.11   & 01:28-02:25 &(-63$^{\prime\prime}$, 67$^{\prime\prime}$)&0$^{\prime\prime}$    &A & 2 &30~s\\
\hline
1999.03.11   & 12:09-13:09 &(-223$^{\prime\prime}$, {-88}$^{\prime\prime}$)&280$^{\prime\prime}$  &A & 2 &30~s\\
\hline
\end{tabular}
\end{table*}

Hydrogen is the most abundant element in the solar atmosphere and
its resonance lines play an important role in the energy transport
of the Sun \citep{Fontenla88}. \lyb~is the second prominent line in
the H Lyman series. Important information on the highly dynamic TR
structures may be carried by the profiles of this line. Early rocket
and satellite observations obtained some \lyb~ profiles
\citep{Reeves76,Lemaire78,Vial82}. However, the profiles obtained in
these early observations suffered from geocoronal absorption.
Theoretical models suggested that the reversal at the center of the
\lyb~ profiles is formed in the upper chromosphere and lower
transition region, while the wings formed in the lower chromosphere
\citep{Gouttebroze78,Barsi79,Schmieder98}. Recently, the
\lyb~profiles obtained with the SUMER instrument have been
extensively investigated. Most \lyb~ profiles appear to have a
non-Gaussian shape with a self-reversal at the center and two peaks
aside, with different shapes in different regions
\citep{Warren98,Heinzel01,Xia03,Xia04,Vial07,Schmieder07,Curdt08,Tian09a,Tian09b,Tian09c,Curdt10a}.
It is believed that the asymmetries of the \lyb~ profiles are
probably caused by the combined effect of flows and opacity in
different layers of the solar atmosphere
\citep{Fontenla02,Gunar08,Tian09b}. Higher-order Lyman line profiles
were also studied. For example, \cite{Warren98} found that the
average profiles for \lyb~through {Ly$\epsilon$\,($n$=5)} are
self-reversed and the remaining lines are flat-topped, and
\cite{Madjarska02} found that profiles through Ly-6 to Ly-11 reveal
self-absorption during EEs. \cite{Madjarska02} suggested that the
observed central depression during EEs in Lyman lines may be mainly
due to an emission increase in the wings.

Although previous studies have demonstrated that hydrogen Lyman
series behave very differently in different solar regions, it is
clear that more data need to be analyzed to advance our knowledge.
As the second prominent line of the hydrogen Lyman series, \lyb~ has
been frequently used in SUMER observations, and can thus provide a
valuable tool to diagnose different structures and properties in
various solar regions.

In this paper, we use co-temporal observations of O~{\sc{vi}},
C~{\sc{ii}}, and \lyb~in a quiet-Sun region (QS) and an equatorial
coronal hole (ECH), to search for signatures of EEs in \lyb~profiles
in these different solar regions. The correlation between the
increased peak emission of \lyb~profiles and the enhanced wing
emission of O~{\sc{vi}} and C~{\sc{ii}} is investigated and
discussed.

\section{OBSERVATIONS AND DATA ANALYSIS}

Information of the SUMER observations is listed in Table~\ref{tab1}.
The first data set was taken in the quiet Sun, and the second one
was obtained in an equatorial coronal hole. The solar X (East-West)
refers to the coordinate range of the scanned region. The solar Y
(South-North) refers to the coordinate of the slit center. Each of
the data includes O~{\sc{vi}} (1031.9 \AA,
$T_e\approx3\times10^5$K), C~{\sc{ii}} (1037.0 \AA,
$T_e\approx5\times10^4$K), and H I \lyb~(1025.7 \AA,
$T_e\approx2\times10^4$K) lines and a series of full detector
readouts at different wavelengths. The scanned regions are outlined
by white rectangles and overlapped on the EIT 195 images (see
Fig.\ref{Fig_1}).

We applied the standard procedures for correcting and calibrating
the SUMER raw data. They include decompression, reversal,
flat-field, dead-time, local-gain, and geometrical corrections. We
extracted the raster scan coordinates from the head-data files of
SUMER and eliminated effects of the solar rotation. The coalignment
of images obtained by different instruments was achieved through a
cross-correlation between the \lyb~intensity maps, the EIT images
and MDI magnetograms.

EEs were identified by O~{\sc{vi}} and C~{\sc{ii}} profiles,
respectively. We first used the procedure described in \cite{Xia03}
to deduce the widths of all spectra and calculated the standard
deviation of the widths. We disregarded the noisy profiles with a
peak intensity smaller than the half-peak intensity of the average
profile. Then the profiles with a width larger than three standard
deviations (3$\sigma$) were singled out for further visual
inspection to finally determine the occurrence of EEs. Our method is
similar to those used by \cite{Teriaca04}.

\section{RESULTS}

\begin{figure*}
   \centering
   \includegraphics[width=17cm]{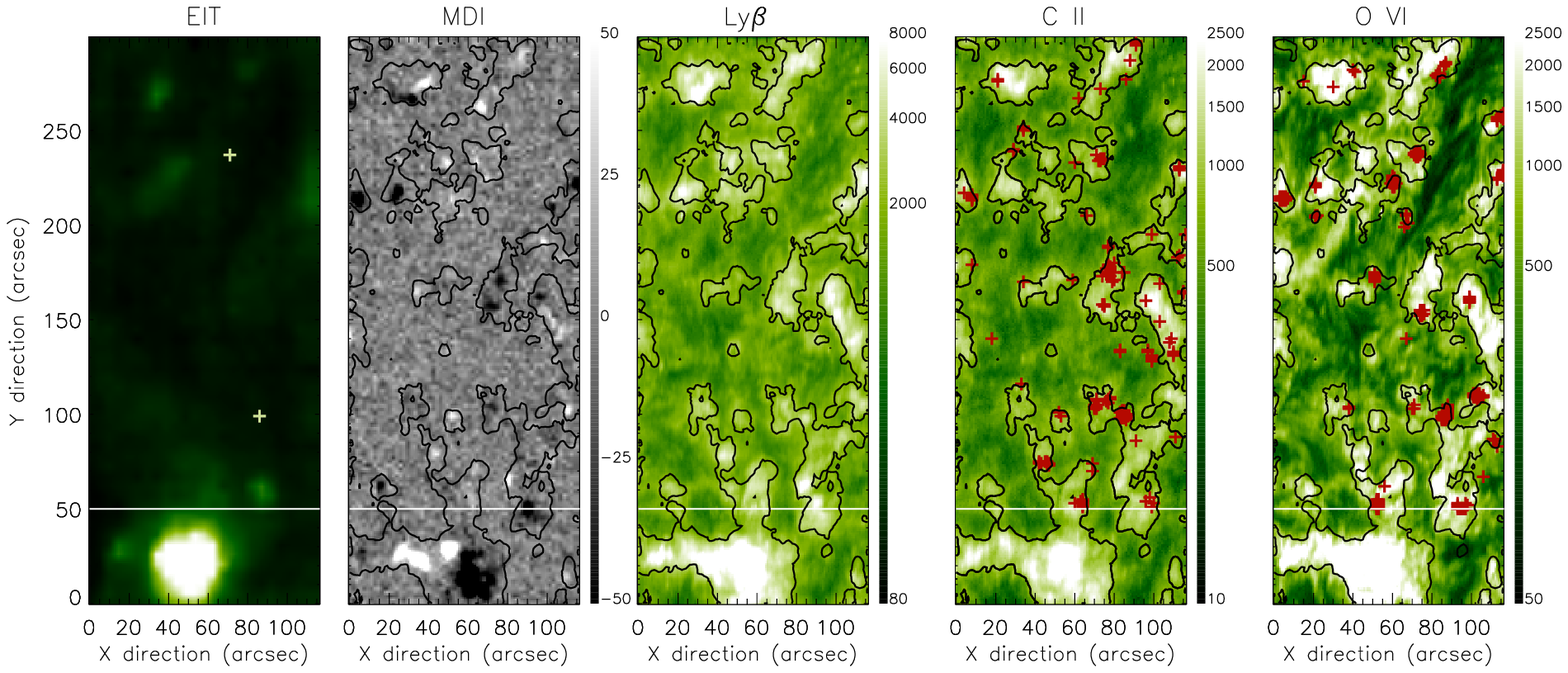}
   \includegraphics[width=17cm]{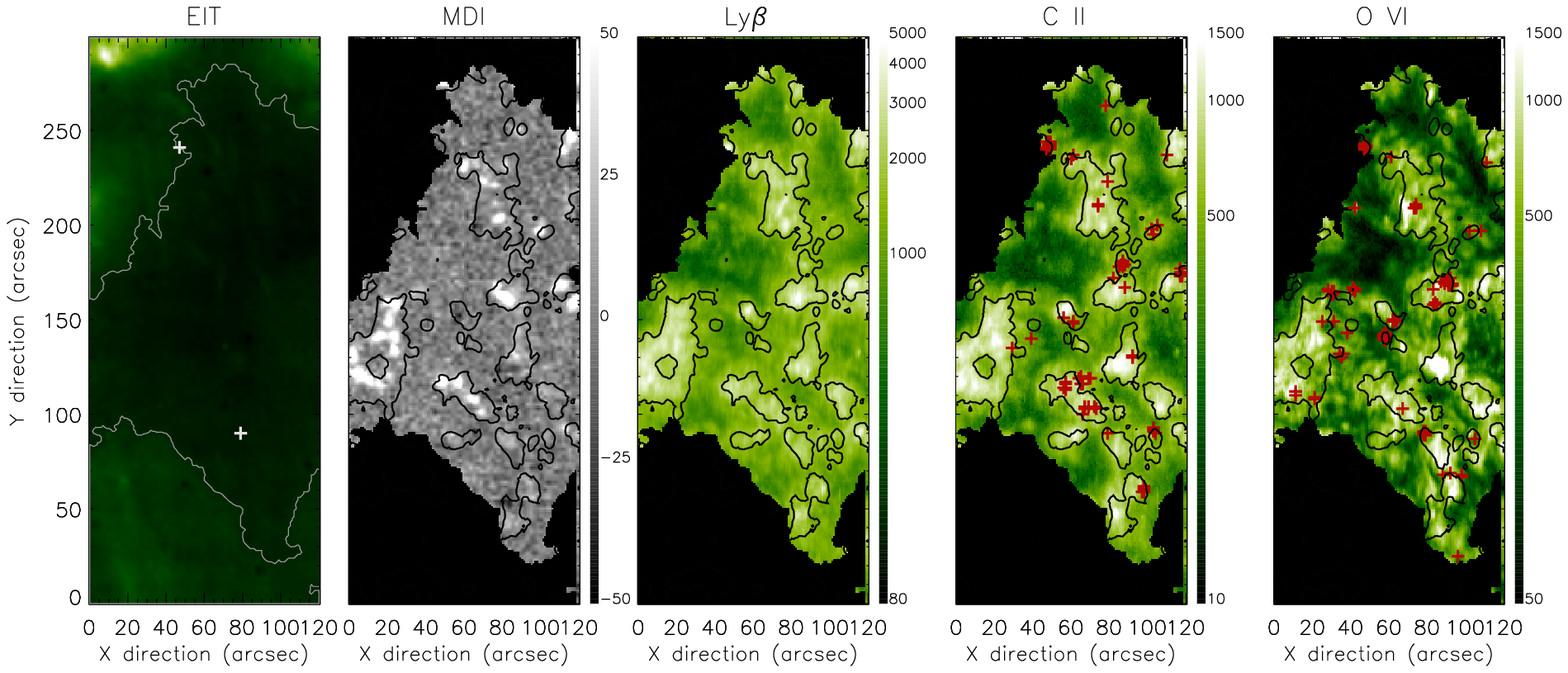}
   \caption{ EIT images in the 195~{\AA} passband, magnetograms (unit: G) obtained by MDI, intensity
   maps (unit: counts/30s/line) of \lyb~, C~{\sc{ii}} and O~{\sc{vi}} (top: QS bottom:
   ECH). Contours overlaid on the magnetogram and intensity maps of SUMER, obtained from the continuum around 1032~\AA,
   outline the chromospheric network. EEs are marked with the red '+' symbols. The '+' symbols in EIT images mark locations
   of four EEs which are analyzed in Fig.\ref{Fig_3}.}\label{Fig_2}
\end{figure*}

Figure~\ref{Fig_2} shows the EIT images in the 195~{\AA} passband,
magnetograms obtained by MDI, intensity maps of \lyb~, C~{\sc{ii}}
and O~{\sc{vi}}. In the upper panel, in order to remove the
interference of a coronal bright point (seen in the bottom of the
EIT map), we used only the region above the horizontal line to
calculate the averaged quiet-Sun profiles. In the lower panel, the
boundary of the equatorial coronal hole was determined with the
intensity threshold of the EIT image \citep{Xia03}. The magnetogram,
intensity maps of \lyb~,C~{\sc{ii}} and O~{\sc{vi}} are overlaid by
black contours to outline the chromospheric network, which occupies
33\% of the whole area and is characterized by the highest
intensities of the continuum around 1032{~\AA}. It is clear that the
network coincides with the concentration of strong photospheric
magnetic fields. In the QS region, strong magnetic fields with
positive (white) and negative (black) polarities are both present
inside the network, while in the ECH region the network regions are
dominated by strong positive magnetic fields and only a few weak
mixed-polarity fields are present. The network structures indicated
by the continuum intensity coincide well with the strong emission of
the three lines. There are many loop-like structures which have
visible footpoints lying on the edge of networks and extend into the
cell interiors. The loop-like structures can be identified much
easier in the ECH than in the QS. A more detailed discussion about
the morphology in these two regions can be found in \cite{Xia04}.

Explosive events are best seen in typical transition-region lines
like Si~{\sc{iv}}($T_e\approx8\times10^4$ K), they can generally be
detected in spectral lines with formation temperatures ranging from
$\sim$10$^{4}$ to 5$\times$10$^{5}$ K
\citep{Madjarska02,Teriaca02,Popescu07}. Here we use two
transition-region lines C~{\sc{ii}} and O~{\sc{vi}}, respectively,
to identify EEs. The identified events are referred to as
``C~{\sc{ii}} EEs'' and ``O~{\sc{vi}} EEs'' hereinafter. In
Fig.~\ref{Fig_2}, the red '+' in intensity maps of C~{\sc{ii}} and
O~{\sc{vi}} mark locations of pixels where EEs were identified. We
find 136 EE pixels detected by the C~{\sc{ii}} line and 167 by the
O~{\sc{vi}} line in the QS, and 70 and 78 correspondingly in the
ECH. Neighboring EE pixels in each spectral line are regarded as
given by a single event. The average occurrence rates of EEs in both
regions are then estimated to be about
1$\times10^{-20}cm^{-2}s^{-1}$, which is comparable to that obtained
by \cite{Teriaca04} in a QS region. It is clear that most of the EEs
lie in the network or on the edge of the network, in line with
previous studies. Furthermore, it is interesting to find that the
pixel positions of the EEs observed in the C~{\sc{ii}} and
O~{\sc{vi}} lines are not spatially overlaid with each other in most
cases. However, this doesn't mean that there is no connection
between these two lines during the events. By inspection of detailed
line profiles, when an EE detected only in one spectral line (i.e.,
with the line width wider than 3$\sigma$), the other one recorded in
the same spectral window at the same location often responses
simultaneously and reveals a significant non-Gaussian profile
although its line width is still smaller than 3$\sigma$. Note that
the formation temperature of the C~{\sc{ii}} line is about 5
$\times10^4$ K which is an order lower than that of the O~{\sc{vi}}
line. This difference of line temperature may result in a different
spectral response to an EE. The response may depend on the height
where an EE occurs. A time delay may also exist in the response of
the high temperature line with respect to the lower temperature
line, if an EE bursts at a lower height.

\begin{figure*}
   \sidecaption
   \includegraphics[width=12.5cm]{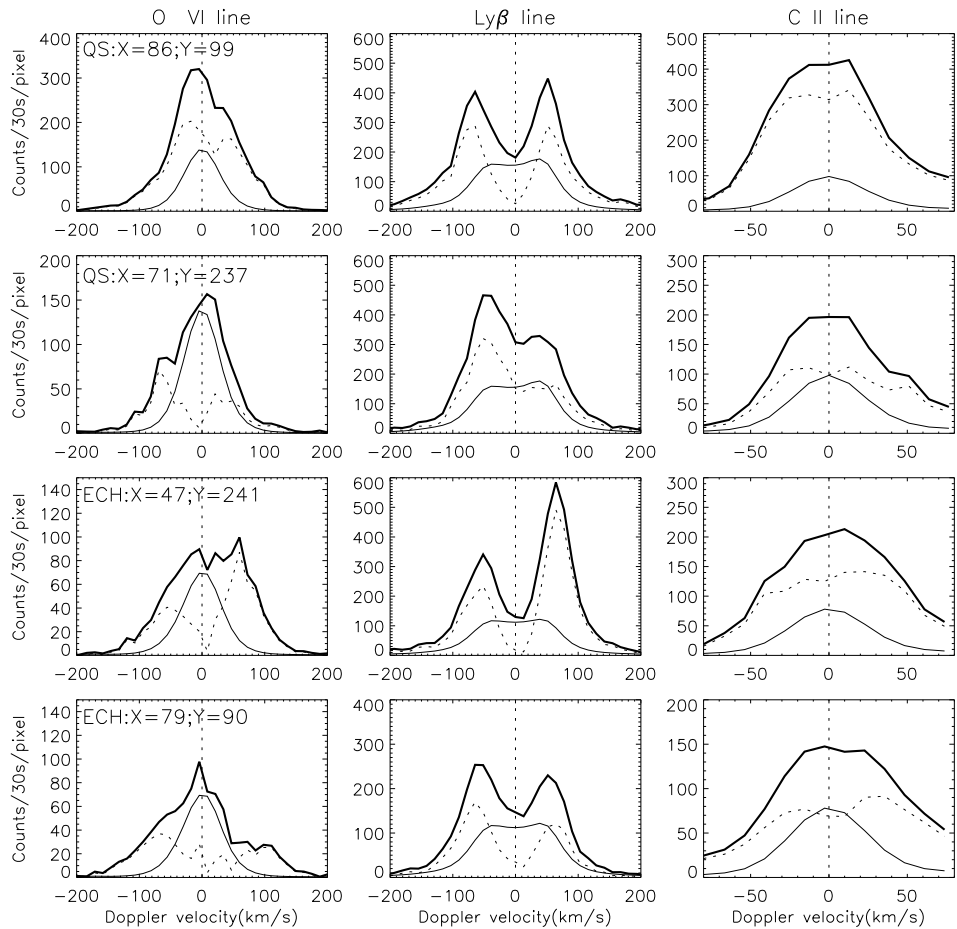}
   \caption{O~{\sc{vi}}, \lyb~, C~{\sc{ii}} profiles of four EEs (marked
   in the EIT images of Fig. 2). Thin lines show the average profiles in the
    QS and ECH. Dotted-lines represent the enhanced radiation by subtracting
    the mean profiles from the EE profiles.}\label{Fig_3}
\end{figure*}

We selected four individual EEs at different locations detected
simultaneously by both the C~{\sc{ii}} and O~{\sc{vi}} lines (two in
QS and two in ECH, two dominated by red peak and two by blue peak).
These EEs are marked on the EIT images shown in Fig.\ref{Fig_2}. In
Fig.\ref{Fig_3}, we present EE profiles of the three lines including
O~{\sc{vi}}, \lyb~and C~{\sc{ii}} (shown by thick lines), as well as
the mean profiles in the whole QS and ECH (shown by thin lines). The
emission enhancements in the wings of the O~{\sc{vi}} and \lyb~lines
are better revealed in the dotted lines which are given by
subtracting the mean profiles from the EE ones. We find that during
the EEs, velocities of the order of 50-100 km/s are clearly present
on the O~{\sc{vi}} line wings, while the C~{\sc{ii}} line presents a
significant bursting feature. Note that we can only plot the
profiles with Doppler velocity of $\pm80$ km/s for the C~{\sc{ii}}
line due to the presence of another two lines (C~{\sc{ii}} at
1036.3\AA~ and O~{\sc{vi}} at 1037.6\AA). It can be seen that the
corresponding \lyb~profiles behave rather differently with a
stronger enhancement at the wings and a deeper reversal at the
center. In most cases, the distance of the two peaks of
\lyb~profiles is apparently larger than that of the mean \lyb~
profiles, and the intensity and position of the wing peaks of the
\lyb~line correlate well with those of the O~{\sc{vi}} line (shown
by dotted lines). For the three events shown in the first, third and
fourth rows of Fig.\ref{Fig_3}, their \lyb~profiles show very small
change of intensity in the line center compared to the mean profile,
although their wings enhance very strongly. Note that the above
descriptions are only for the four selected individual events
detected simultaneously with both the O~{\sc{vi}} and C~{\sc{ii}}
lines. The more general properties of the observed events will be
analyzed below.

\begin{figure*}
\centering
   \includegraphics[width=15cm]{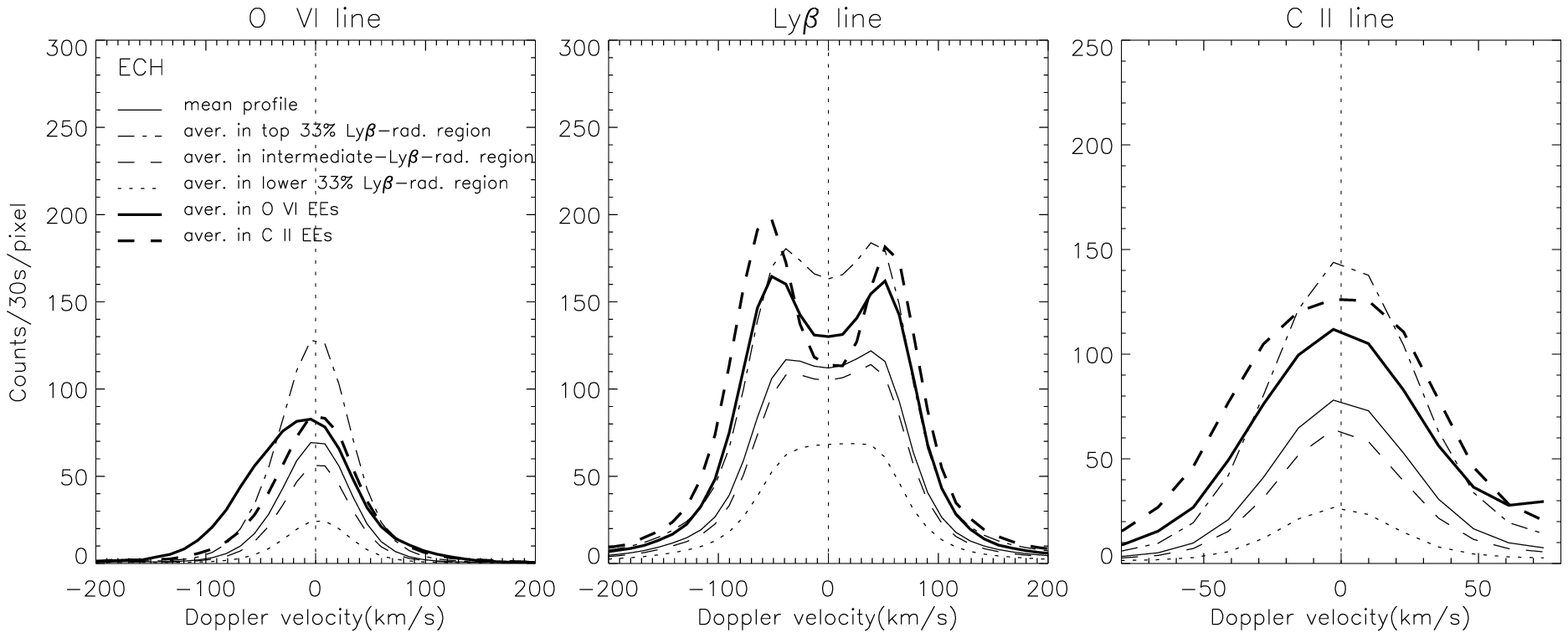}
   \includegraphics[width=15cm]{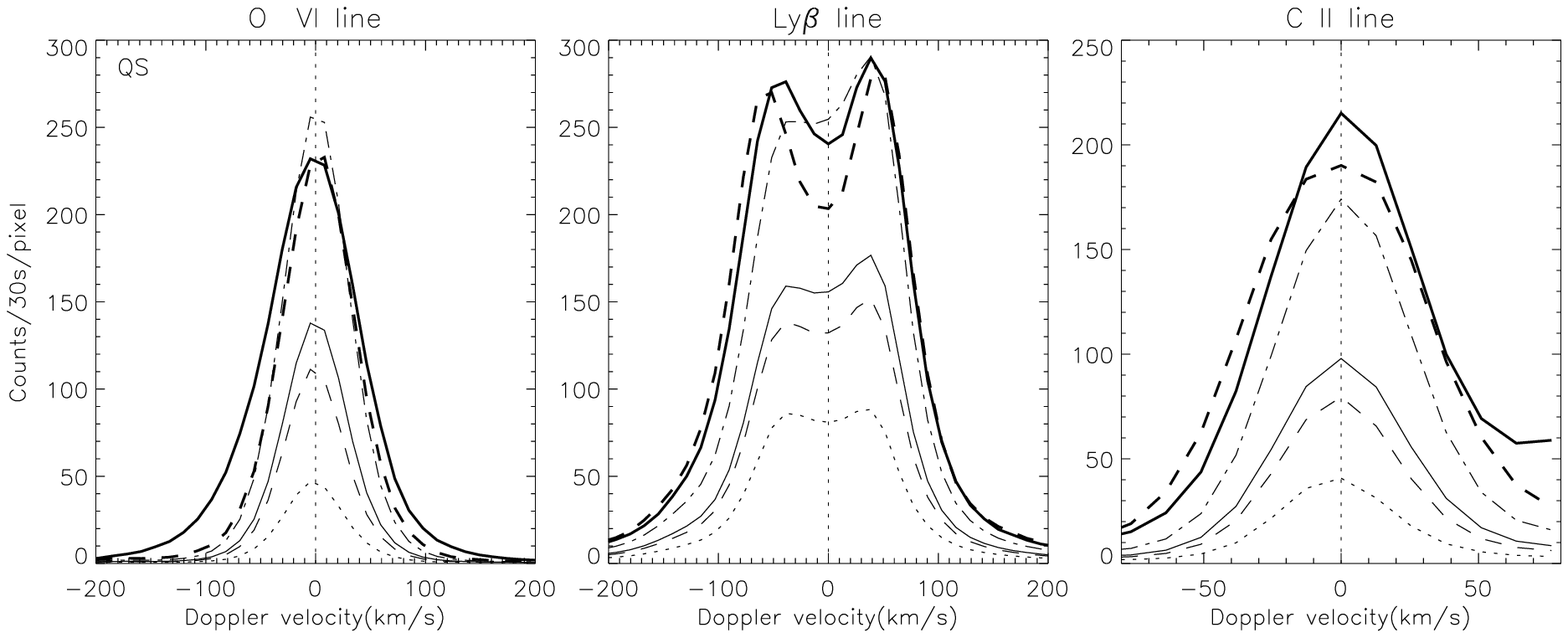}
   \caption{Different kinds of average O~{\sc{vi}}, \lyb~ and C~{\sc{ii}} profiles (left column: O~{\sc{vi}}, middle column: \lyb, right column: C~{\sc{ii}};
   top panels: ECH, bottom panels: QS). Thin solid line is the average profile of each region of entire ECH and QS.
    Thin dash-dotted, thin dotted and thin dashed lines represent profiles averaged in the top 33\%, lower 33\%, and
    intermediate-\lyb-radiation regions, respectively. Thick solid line shows the average profile of EEs detected by the O~{\sc{vi}} line and
    thick dashed line by the C~{\sc{ii}} line. See also the legend in the top-left panel.}\label{Fig_4}
\end{figure*}

Figure~\ref{Fig_4} shows different kinds of average O~{\sc{vi}},
\lyb~and C~{\sc{ii}} profiles observed in the QS and ECH regions.
The dashed-vertical lines in each panel indicate the central
position of the profile averaged in the relevant QS or ECH region.
According to the intensity of the \lyb~line, we divided each region
of ECH and QS into three parts: top 33\%, lower 33\%, and
intermediate-radiation regions. Then we calculated the average
O~{\sc{vi}}, \lyb~and C~{\sc{ii}} profiles in each radiation region.
We find that the red peak of \lyb~ profile is higher than the blue
peak in the QS, and the trend becomes more apparent with increasing
intensity of \lyb (seen in bottom panels). In the ECH, the self
reversal at the center of the \lyb~ profile is obvious and a deeper
one is observed with increasing intensity, while the strengths of
two peaks are basically the same (seen in top panels).

In Fig.~\ref{Fig_4}, we also plot the average O~{\sc{vi}}, \lyb~and
C~{\sc{ii}} profiles of the O~{\sc{vi}} EEs (shown by thick solid
lines) and the C~{\sc{ii}} ones (shown by thick dashed lines),
respectively. It can be seen that the average \lyb~profiles of the
EEs in both ECH and QS regions show a deeper self-reversal and two
prominent wing peaks, and the trend is more obvious for the
C~{\sc{ii}} EEs than the O~{\sc{vi}} ones. In the ECH, compared with
the mean ECH profile(shown by thin solid lines), the average
O~{\sc{vi}} profile of the O~{\sc{vi}} EEs has a broader width and
is shifted towards the blue side, while that of the C~{\sc{ii}} EEs
is not very different. The C~{\sc{ii}} line of the O~{\sc{vi}} EEs
is on average broader than that of the mean ECH profile, and that of
the C~{\sc{ii}} EEs is even broader. They both tend to have a more
enhanced blue wing. And again, the blue wing of the C~{\sc{ii}} EEs
is more enhanced than that of the O~{\sc{vi}} ones. For the \lyb~
line, the average profile of the O~{\sc{vi}} EEs is almost symmetric
and that of the C~{\sc{ii}} EEs has an obviously stronger blue peak.
The distances of the two peaks observed in both the O~{\sc{vi}} EEs
and C~{\sc{ii}} EEs are larger than that of the mean ECH profile,
and that for the C~{\sc{ii}} EEs is the largest. In the QS, similar
trends can be found for the widths of the C~{\sc{ii}} and
O~{\sc{vi}} lines. However, the C~{\sc{ii}} profile of the
O~{\sc{vi}} EEs shows a more enhanced red wing, which may be at
least partly caused by the greatly enhanced blue wing of another
O~{\sc{vi}} line at 1037.6\AA. And, for the \lyb~line, the red peak
is stronger in the QS, in contrast with the features observed in the
ECH. The distance of the two peaks in the QS also shows a similar
trend as that in the ECH.

\begin{figure*}
   \includegraphics[width=17cm]{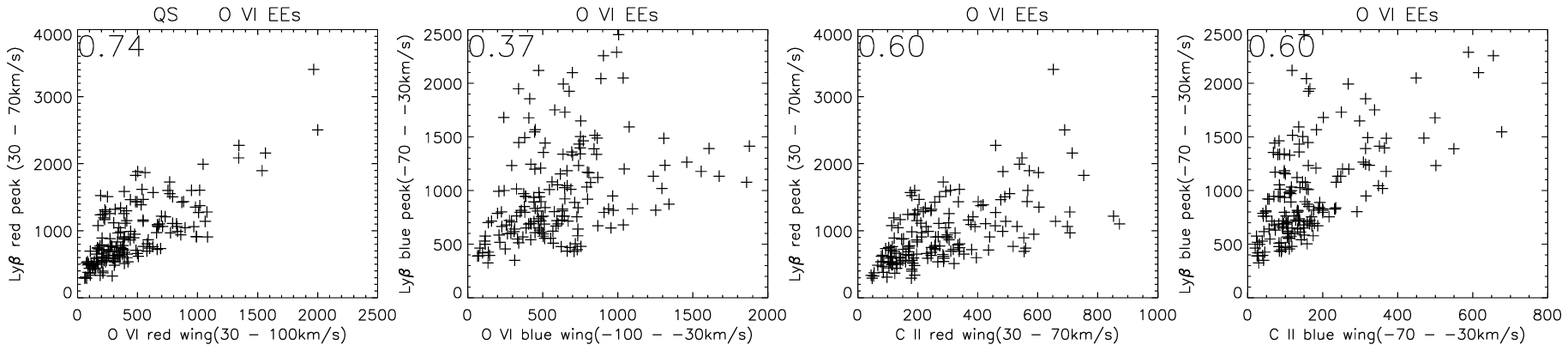}
   \includegraphics[width=17cm]{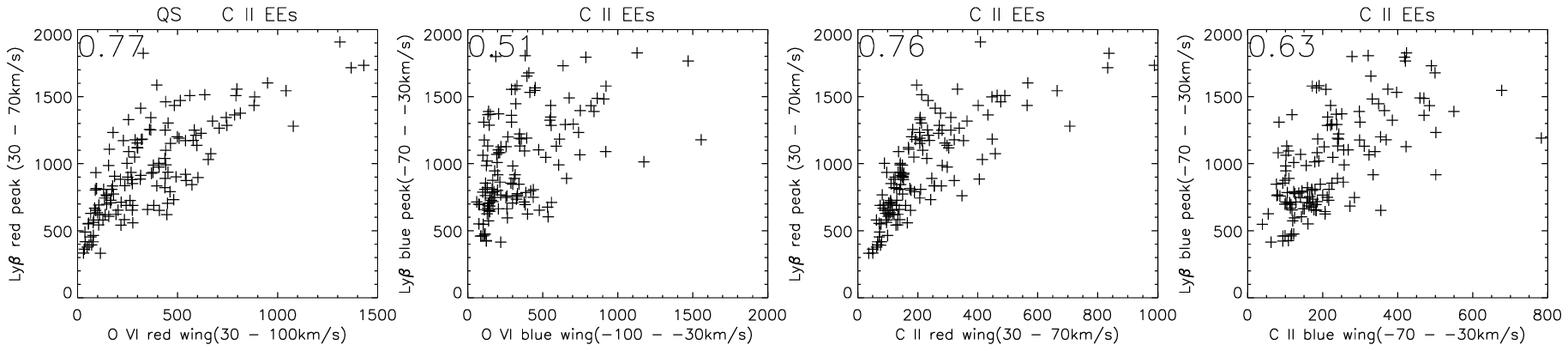}
   \includegraphics[width=17cm]{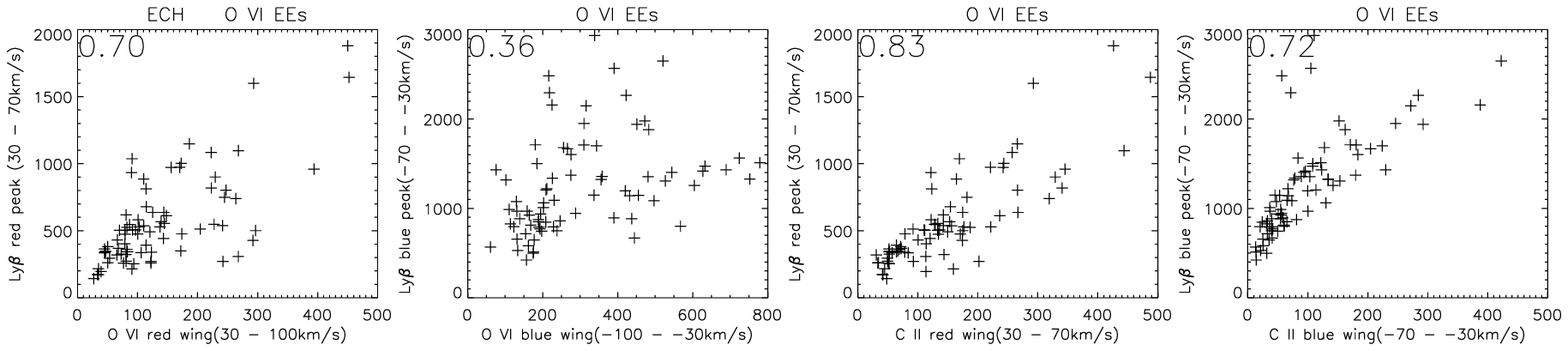}
   \includegraphics[width=17cm]{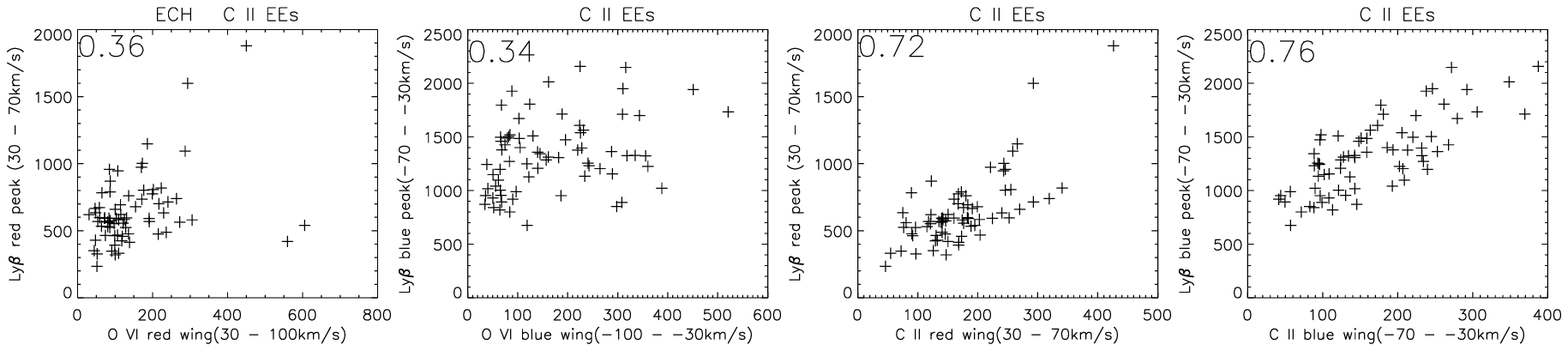}
   \caption{Relationship between photon counts of blue/red wing
    of O~{\sc{vi}} and C~{\sc{ii}} profiles and photon counts of blue/red peak of \lyb~profiles during EEs.
    The correlation coefficient is shown in each panel (upper two panels: QS; bottom two panels: ECH) }\label{Fig_5}
\end{figure*}

\begin{table}
  \caption[]{Correlation coefficients of enhanced emission during EEs}
  \label{tab2} \centering
  \begin{tabular}{ccccc}
  \hline 
correlated parameters&\multicolumn{2}{c}{O~{\sc{vi}} EEs}&\multicolumn{2}{c}{C~{\sc{ii}} EEs}\\
\hline
 & red & blue & red & blue  \\
\hline
QS (O~{\sc{vi}} wing $\sim$ \lyb~ peak)&0.74&0.37&0.77&0.51\\
\hline
QS (C~{\sc{ii}} wing $\sim$ \lyb~ peak)&0.60&0.60&0.76&0.63\\
\hline
CH (O~{\sc{vi}} wing $\sim$ \lyb~ peak)&0.70&0.36&0.36&0.34\\
\hline
CH (C~{\sc{ii}} wing $\sim$ \lyb~ peak)&0.83&0.72&0.72&0.76\\
\hline
\end{tabular}
\end{table}

In order to quantify the correlation between the increased peak
emission of \lyb~profiles and the enhanced wing emission of
O~{\sc{vi}}, we calculated the photon counts of blue/red wing
(Doppler velocity from 30 km/s to 100 km/s) of O~{\sc{vi}} profiles
and the photon counts of blue/red peak (Doppler velocity from 30
km/s to 70 km/s ) of \lyb~profiles at EE pixels in the QS and ECH,
respectively. In the same way, we also calculated the correlation
coefficients between the C~{\sc{ii}} wings (Doppler velocity from 30
km/s to 70 km/s) and the \lyb~peaks. Figure~\ref{Fig_5} presents the
corresponding scatter plots. We also list the calculated correlation
coefficients in Table~\ref{tab2}, which are all positive. It seems
that the enhancement of the \lyb~peaks represents the signature of
EEs. Furthermore, the correlation seems to be quite good for all
red/blue wings of C~{\sc{ii}} profiles and all red wings of
O~{\sc{vi}} profiles. For the O~{\sc{vi}} line, the correlation
seems to be weaker on the blue than the red side. Note that the
formation temperature of the C~{\sc{ii}} line is much closer to that
of the \lyb~line than that of the O~{\sc{vi}} line. This may explain
the better correlation between the increased peak emission of the
\lyb~line and the enhanced C~{\sc{ii}} wings during EEs.

\section{DISCUSSION}

The major finding of this paper is that there is a clear correlation
between the increased peak emission of \lyb~profiles and the
enhanced wing emission of the transition-region lines, especially
the C~{\sc{ii}} line, which has a formation temperature close to
that of the \lyb~ line. This result indicates that EEs can greatly
modify \lyb~profiles, especially the two peaks of the profiles. The
clear correlation suggests that EEs are responsible for the enhanced
peak emission of \lyb.

We can assume that the \lyb~emission during EEs has two components,
the background emission and the jet emission. The former is the
emission from the background QS or CH. Its source lies in the upper
chromosphere and lower TR. As it propagates to the upper atmosphere,
emission from the central part of the profile is absorbed by the
atomic hydrogen, revealing a central depression in the profile. On
the other hand, the jet emission is largely different. Jets produced
by EEs can heat the relatively cold background plasma causing
enhanced ionization and further emission in the whole profile of
colder lines. This is confirmed by the jet emission of C~{\sc{ii}}
shown in Fig.\ref{Fig_3}. At the same time, the plasma can also be
accelerated to a much higher velocity causing greatly enhanced
emission in their line wings. Since the jets are usually
bidirectional with a high speed, the \lyb~photons emitted by the
jets should also be Doppler-shifted towards both longer and shorter
wavelengths. If the speed of the jets has a line-of-sight component,
we should observe these Doppler-shifted \lyb~emission which is added
to the almost-at-rest background \lyb~emission, causing enhancement
of the peaks of the background \lyb~profiles. The jet-emitted
\lyb~profiles experience much less radiative transfer process. This
is because that the EEs are most prominent in the middle and upper
TR, above which the density is very low and the atomic hydrogen can
not significantly absorb the emission from below. Also, the almost
rest coronal atmosphere could not absorb the Doppler-shifted jet
emission due to the lack of the wavelength match.

\cite{Madjarska02} found that profiles through Ly-6 to Ly-11 reveal
self-absorption during EEs. The authors concluded that the observed
central depression during EEs in Lyman lines may be mainly due to an
emission increase in the wings. Our analysis of the \lyb~profiles
during EEs suggests that the jets produced by EEs emit
Doppler-shifted \lyb~photons and cause enhanced emission at the
peaks of \lyb~profiles. Our result complements that in
\cite{Madjarska02}. In addition, most previous studies on EE-like
dynamic events were conducted based on analysis of optically-thin
spectral lines (such as Si IV and O VI lines). Our result further
indicates that \lyb~and other Lyman lines could be used to identify
these transient events even in absence of strong spectral lines in
the transition region. Since \lyb~is the second prominent line in
the hydrogen Lyman series and is much more frequently used in
observations, the variation of the \lyb~profiles provides a good
tool to diagnose different structures and properties in different
regions. Our finding of the signatures of EEs in \lyb~profiles is
thus helpful to investigate the thermodynamics of the jets produced
by EEs.

The average \lyb~profiles of EEs have an obviously stronger red peak
in the QS, while in the ECH the blue peak seems to be stronger for
the C~{\sc{ii}} EEs. The different relative strengths of the
blue-shifted and red-shifted jet-components might account for the
different asymmetries. In Section 3, we have discussed the average
line widths of EE pixels and found the C~{\sc{ii}} profile of the
C~{\sc{ii}} EEs in the ECH has an enhanced blue wing being more
pronounced than other profiles. Correspondingly, the blue peak of
\lyb~is relatively stronger and the peak separation is larger. The
blue-shift of EEs may cause the relatively significant blue peak of
\lyb~profiles. As we know, fast bidirectional jets can lead to the
separation of the two peaks. From line profiles of the four typical
EEs, we suggest that the different asymmetry of \lyb~profiles seems
to be a result of different speed and strength of EEs' jets.

In the quiet Sun, most \lyb~profiles are found to have a stronger
red peak \citep{Warren98}. In fact, the \lyb~profile has different
shapes in different regions. \citet{Xia03} and \citet{Xia04} found
that there are more \lyb~profiles with a stronger blue peak in
equatorial coronal holes than in the quiet Sun, so that the red-peak
asymmetry of the average \lyb~profile is less pronounced in the ECH.
\citet{Tian09b} found that \lyb~profiles in polar coronal hole have
a stronger blue-peak which is opposite to those in the QS. Here we
find that the average \lyb~profile in the ECH has almost symmetrical
peaks. The different asymmetries of the \lyb~profiles might reflect
different flow fields of the upper solar atmosphere in different
parts of the Sun. The most prominent difference of systematic flow
systems between the polar coronal hole and quiet-Sun regions is that
upflows are predominant in the upper TR of polar coronal holes
\citep{Dammasch99,Hassler99,Tu05,Tian10}, while upflows are
localized at network junctions in the upper TR of the quiet Sun
\citep{Hassler99,Tian08,Tian09d}. In ECHs, the flow pattern in the
upper TR might be similar to that of the polar coronal hole but the
magnitude of the upflows might be smaller
\citep{XiaEtal03,Aiouaz05,Raju09}. So the average \lyb~profile in
the ECH reveals an almost symmetrical shape, in an intermediate
phase between the red-peak-dominance in the quiet Sun and
blue-peak-dominance in polar coronal holes. Another possibility
might come from the larger opacity in the coronal hole. Our data
reveals that the \lyb~profiles are on average more reversed in the
the ECH than in the QS, which indicates that the opacity is larger
in the ECH. This finding complements the previous finding of a
larger opacity in polar coronal holes than in the quiet Sun
\citep{Tian09b}. We may assume that the \lyb~line behaves more or
less similar to typical TR lines in the quiet Sun. In polar coronal
holes, the opacity is so large that the \lyb~line now behaves more
similar to \lya~, with a stronger blue peak \citep{Curdt10b}. If the
opacity in the ECH is larger than that in the QS but smaller than
that in the polar coronal holes, then it is not surprising to
observe the almost symmetrical \lyb~profile in the ECH.

The average and four typical \lyb~profiles imply the contribution
from the wings of EEs. We qualitatively analyzed the relevance
between the wings of EEs and peaks of \lyb~profiles. The relevance
between C~{\sc{ii}} wings of EEs and \lyb~peaks is very significant,
but for the O~{\sc{vi}} line, the relevance between blue wings of
EEs and blue peaks is genenally poorer than that between their red
counterparts, especially in the ECH. As we know, C~{\sc{ii}} has a
formation temperature close to that of \lyb, which is about an order
lower than that of O~{\sc{vi}}. During EEs, it seems that in
general, the variation of \lyb~peaks correlates more closely with
that of C~{\sc{ii}}. However, the scan data used here have no
information of time. A larger error appears if the time difference
exists among the different lines which are emitted in different
layers of the solar atmosphere when EE occurs, especially when we
are using the scan data with larger exposure time. For example,
\cite{Madjarska02} have observed a time delay of about 20-40 s in
the response of the transition-region line (S VI, 200 000K) with
respect to the chromospheric line (H I Ly 6, 20 000K), when an EE
can be seen in a chromospheric line. Therefore, this conclusion
should be verified in the future by analyzing more data, especially
time-series data with short exposure times.

Previous studies have confirmed that the line shapes of \lya~and
\lyb~were significantly affected by quasi-steady flow field in the
transition region \citep{Curdt08,Tian09a}. In this paper, we find
the \lyb~profiles are also modified by the transient flow field
generated by EEs in QS and ECH. This finding implies that one should
be careful in the modelling and interpretation of such observational
data. According to our results, Lyman profiles, especially when
observed at a high spatial and temporal resolution, are affected by
both EEs and opacity. When the underlying dynamic process of the
solar atmosphere is analyzed by using \lyb~and other Lyman lines,
one should consider not only the line source function and opacity,
but also the flow field in the transition region, including both the
quasi-steady and transient flows. For instance, one needs to take
all these factors into account for the numerical simulation in order
to explain the observed line shapes of Lyman series and their
relation with the flow field.

The \lyb~line is the second strongest line of hydrogen Lyman series.
Some observational features of this line are similar to those of
\lya, while some are very different. The opacity of \lya~is much
larger than that of \lyb. It is interesting to ask whether there are
similar behaviours of \lya~profile during EEs, which needs to be
addressed in the future. Since hydrogen is the most abundant
component of the Sun and \lya~is the most prominent line emitted by
the chromosphere and lower transition region, such studies could
thus be important for the future high-resolution observations of
Lyman lines. Moreover, it is also interesting to look for signatures
of other solar dynamic events (such as flares and CMEs) in Lyman
lines in order to study these dramatic eruptions. As the formation
height of Lyman lines in the solar atmosphere is relatively low,
their response to the events could be used to study the initiations
of these eruptions and to advance the predicting technology of the
associated space weather events.

\section{CONCLUSION}

We have used co-temporal observations of O~{\sc{vi}}, C~{\sc{ii}}
and \lyb~in a quiet-Sun region and an equatorial coronal hole to
search for signatures of explosive events in \lyb~profiles. We find
that EEs have significant impacts on the profiles of \lyb. During
EEs, the center of \lyb~ profiles becomes more reversed and the
distance of the two peaks becomes larger, both in the equational
coronal hole and in the quiet Sun. The average \lyb~ profile of the
EEs detected by C~{\sc{ii}} has an obvious stronger blue peak.
Statistical analysis shows that there is a clear correlation between
the increased peak emission of \lyb~profiles and the enhanced wing
emission of C~{\sc{ii}} and O~{\sc{vi}}. The correlation is more
obvious for the \lyb~ peaks and C~{\sc{ii}} wings, and less
significant for the \lyb~ blue peak and O~{\sc{vi}} blue wing. It
indicates that the jets produced by EEs emit Doppler-shifted
\lyb~photons, causing enhanced emission at positions of the peaks of
\lyb~profiles. The more-reversed \lyb~ profiles confirm the presence
of a larger opacity in the coronal hole than in the quiet Sun.

\begin{acknowledgements}
The authors thank the referee for his/her comments and suggestions
that improved the manuscript. The SUMER project is financially
supported by DLR, CNES, NASA, and the ESA PRODEX Programme (Swiss
contribution). SUMER is an instrument onboard {\it SOHO}, a mission
operated by ESA and NASA. We thank Dr. W. Curdt for the helpful
discussions. The work is supported by the National Natural Science
Foundation of China(NSFC) under contracts 40974105 and 40774080,
40890162, and NSBRSF G2006CB806304 in China.
\end{acknowledgements}

\end{document}